\documentclass[iop]{emulateapj-rtx4}
\usepackage{natbib}
\usepackage{color}
\usepackage{graphicx}
\newcommand{\Ka}{K$\alpha$}

\newcommand{\Msun}{$M_{\odot}$}

\shorttitle{Fe-K Emission from Supernova Remnants}
\shortauthors{Yamaguchi et al.}

\begin{document}

\title{Discriminating the Progenitor Type of Supernova Remnants\\ 
with Iron K-Shell Emission}

\author{Hiroya Yamaguchi\altaffilmark{1,2,3},
Carles Badenes\altaffilmark{4}, 
Robert Petre\altaffilmark{1},
Toshio Nakano\altaffilmark{5},
Daniel Castro\altaffilmark{6},\\
Teruaki Enoto\altaffilmark{1,7},
Junko S.\ Hiraga\altaffilmark{5},
John P.\ Hughes\altaffilmark{8},
Yoshitomo Maeda\altaffilmark{9},
Masayoshi Nobukawa\altaffilmark{10},\\
Samar Safi-Harb\altaffilmark{11,12},
Patrick O.\ Slane\altaffilmark{3},
Randall K.\ Smith\altaffilmark{3},
Hiroyuki Uchida\altaffilmark{10}
}
\email{hiroya.yamaguchi@nasa.gov}

\altaffiltext{1}{NASA Goddard Space Flight Center, Code 662, Greenbelt, MD 20771, USA}
\altaffiltext{2}{Department of Astronomy, University of Maryland, College Park, MD 20742, USA}
\altaffiltext{3}{Harvard-Smithsonian Center for Astrophysics, 60 Garden Street, 
	Cambridge, MA 02138, USA}
\altaffiltext{4}{Department of Physics and Astronomy and Pittsburgh Particle Physics, 
	Astrophysics and Cosmology Center (PITT PACC), University of Pittsburgh, 
	3941 O'Hara St, Pittsburgh, PA 15260, USA}
\altaffiltext{5}{Department of Physics, The University of Tokyo, Bunkyo, Tokyo, 113-0033, Japan}
\altaffiltext{6}{MIT-Kavli Center for Astrophysics and Space Research, 77 Massachusetts Avenue, 
	Cambridge, MA 02139, USA}
\altaffiltext{7}{RIKEN (The Institute of Physical and Chemical Research), 2-1 Hirosawa, Wako, 
	Saitama 351-0198, Japan}
\altaffiltext{8}{Department of Physics and Astronomy, Rutgers University, 
	136 Frelinghuysen Road, Piscataway, NJ 08854, USA}
\altaffiltext{9}{Institute of Space and Astronautical Science, JAXA, 3-1-1 Yoshinodai, Sagamihara, 
	Kanagawa 229-8510, Japan}
\altaffiltext{10}{Department of Physics, Kyoto University, Kitashirakawa-oiwake-cho, Sakyo-ku, 
	Kyoto 606-8502, Japan}
\altaffiltext{11}{Department of Physics and Astronomy, University of Manitoba, Winnipeg, 
	MB R3T 2N2, Canada}
\altaffiltext{12}{Canada Research Chair}

\begin{abstract}

Supernova remnants (SNRs) retain crucial information about both their parent 
explosion and circumstellar material left behind by their progenitor. 
However, the complexity of the interaction between supernova ejecta and ambient 
medium often blurs this information, and it is not uncommon for the basic progenitor 
type (Ia or core-collapse) of well-studied remnants to remain uncertain. 
Here we present a powerful new observational diagnostic to discriminate between 
progenitor types and constrain the ambient medium density of SNRs solely using 
Fe K-shell X-ray emission. We analyze all extant {\it Suzaku} observations of SNRs and 
detect Fe K$\alpha$ emission from 23 young or middle-aged remnants, including five 
first detections (IC\,443, G292.0+1.8, G337.2--0.7, N49, and N63A). 
The Fe K$\alpha$ centroids clearly separate progenitor types, with the Fe-rich ejecta 
in Type Ia remnants being significantly less ionized than in core-collapse SNRs. 
Within each progenitor group, the Fe K$\alpha$ luminosity and centroid are well correlated, 
with more luminous objects having more highly ionized Fe. Our results indicate that 
there is a strong connection between explosion type and ambient medium density, 
and suggest that Type Ia supernova progenitors do not substantially modify their 
surroundings at radii of up to several parsecs. 
We also detect a K-shell radiative recombination continuum of Fe 
in W49B and IC\,443, implying a strong circumstellar interaction in the early evolutionary 
phases of these core-collapse remnants. 

\end{abstract}


\section{Introduction}
\label{sec:intro}

Supernova remnants (SNRs) provide unique insights into both the supernova (SN) explosion 
that generated them and the ambient medium that 
surrounded their progenitors at the time of the explosion. 
Unfortunately, the complex physical processes involved in the interaction 
between ejecta and ambient medium often blur this information, to the point that the explosion type 
(i.e., Type Ia or core-collapse: Ia and CC hereafter) of several well-studied 
SNRs still remains controversial.

The X-ray emission from young and middle-aged SNRs is ideally suited to disentangle 
the contributions from the SN explosion and circumstellar interaction 
\citep[see][for a recent review]{Vink12}. 
Their thermal X-ray spectra are often dominated by strong optically-thin emission lines 
from ejecta that retain the nucleosynthetic signature of their birth events. 
On the other hand, the X-ray emitting plasma is in a state of non-equilibrium ionization (NEI), 
and its time-dependent ionization degree is controlled by the ambient medium density, 
which is a sensitive diagnostic of the presence of circumstellar material (CSM) left behind by 
the SN progenitor \citep[e.g.,][]{Badenes05,Badenes07}.

Indeed, much progress has been made in the typing of SNRs using their X-ray emission. 
Using {\it ASCA} data, \cite{Hughes95} showed that it is possible to distinguish Ia remnants 
from CC ones by virtue of their ejecta composition; Fe-rich and O-poor SNRs are likely Ia, 
while SNRs dominated by O and Ne lines with weak Fe L emission are likely CC. 
More recently, \cite{Lopez09b,Lopez11} argued that {\it Chandra} images of Ia SNRs 
show a higher degree of symmetry than those of CC SNRs. This result implies that 
CC SNe are more asymmetric than Ia SNe, and/or CC SNRs expand into more 
asymmetric CSM. These methods are promising, but require sophisticated analysis 
techniques whose results might lead to ambiguous interpretations. 
Abundance determination in NEI plasmas is notoriously uncertain 
\citep[see][for a discussion]{Borkowski01}, and neither of these methods easily leads to 
placement of quantitative constraints on the presence of CSM in a SNR. 
In this {\it Letter}, we present a new, straightforward observational diagnostic for typing SNRs 
in X-rays that relies only on the centroid and flux of a single spectral line -- 
the Fe K$\alpha$ emission at 6.4--6.7\,keV.

The Fe K line blend is well separated from emission lines of
other abundant elements. 
Since the production of Fe occurs at the heart of an SN explosion, reverse shock 
heating of this element can be delayed compared to elements synthesized in 
the outer layers. This 
often results in an ionization state lower than He-like  (Fe$^{24+}$) in young or middle-aged SNRs.
The ionization state in turn determines the Fe K$\alpha$ centroid \citep[e.g.,][]{Yamaguchi14}, 
which is easily measured using current CCD instruments. 
Furthermore, the Fe K emission is largely unaffected by foreground extinction, 
unlike Fe L-shell blends. 
These spectral advantages and simplicities make our method more straightforward than 
the existing ones, and especially attractive for current and future X-ray missions with 
high throughput, like {\it Suzaku}, {\it XMM-Newton}, and {\it Astro-H}. 
Here we show that the Fe K$\alpha$ centroids (hence the Fe ionization state) 
clearly discriminate the progenitor type and place strong limits on the presence of 
CSM in SNRs at radii of several parsecs, which has important consequences for 
SN progenitor studies.

\section{Data Analysis}
\label{sec:analysis}

\begin{table*}
\begin{center}
\caption{List of the SNRs where Fe-K$\alpha$ emission is detected.$\!^a$
  \label{tab}}
  \begin{tabular}{lcccccccccc}
\hline \hline
Name$^b$ & Obs.\ ID & Exposure & Energy & Photon Flux & $N_{\rm H}$$^c$ & Distance & Radius & Age & BGD$^d$ & Refs. \\
~ & & (ks) & (eV) & {\tiny ($10^{-5}$\,cm$^{-2}$\,s$^{-1}$)} & {\tiny ($10^{22}$\,cm$^{-2}$)} & (kpc) & (pc) & (yr) & &  \\
\hline
\multicolumn{11}{c}{Type Ia SNRs and candidates} \\
\hline
Kepler & 5050920[1--7]0 & 574 & $6438 \pm 1$ & $34.6 \pm 0.2$ & 0.5 & 4.8 & 2.4 & 410 & (1) & 1 \\
3C\,397$^{\dagger}$ & 505008010 & 69 & $6556_{-3}^{+4}$ & $13.7 \pm 0.4 $ & 3.0 & 10.3 & 10.5 & 1500--5500 & (2) & 2,3 \\
Tycho$^{\ast}$ & 5030850[1,2]0 & 416 & $6431 \pm 1$ & $61.0 \pm 0.4$ & 0.7 & 2.8 & 3.4 & 442 & (2) & 4,5 \\
RCW\,86$^{\dagger}$ & (See Note) & 378 & $6408_{-5}^{+4}$ & $14.0 \pm 0.7$ & 0.3 & 2.5 & 16 & 1829 & (3) & 6 \\
SN\,1006$^{\ast}$ & (See Note) & 317 & $6429 \pm 10$ & $2.55 \pm0.43$ & 0.07 & 2.2 & 10 & 1008 & (3) & 7 \\
G337.2--0.7 & 507068010 & 304 & $6505_{-31}^{+26}$ & $0.21 \pm 0.06$ & 4.0 & 9.3 & 8.1 & 5000--7000 & (2) & 8 \\
G344.7--0.1$^{\dagger}$ & 501011010 & 42 & $6463_{-10}^{+9}$ & $4.03 \pm 0.33$ & 5.0 & 14 & 16 & 3000--6000 & (2) & 9 \\
G352.7--0.1$^{\dagger}$ & 506052010 & 202 & $6443_{-12}^{+8}$ & $0.82 \pm 0.08$ & 2.6 & 7.5 & 6.0 & $\sim$5000 & (2) & 10 \\
N103B$^{\dagger}$ & 804039010 & 224 & $6545 \pm 6$ & $2.15 \pm 0.10$ & 0.06 & 50 & 3.6 & $\sim$860 & (2) & 11,12 \\
0509--67.5$^{\ast}$ & 5080720[1,2]0 & 329 & $6425_{-15}^{+14}$ & $0.32 \pm 0.04$ & 0.05 & 50 & 3.6 & $\sim$400 & (2) & 12,13 \\
0519--69.0$^{\ast}$ & 806026010 & 348 & $6498_{-8}^{+6}$ & $0.93 \pm 0.05$ & 0.06 & 50 & 4.0 & $\sim$600 & (2) & 12,14 \\
\hline
\multicolumn{11}{c}{Core-collapse SNRs and candidates} \\
\hline
Sgr\,A~East$^{\ast}$ & (See Note) & 88 & $6664 \pm 3$ & $22.3 \pm 1.0$ & 10 & 8.5 & 3.7 & $\sim$4000 & (2) & 15 \\ 
G0.61+0.01$^{\dagger}$ & 100037060 & 77 & $6634_{-12}^{+14}$ & $3.3 \pm 0.5$ & 16 & 8.5 & 5.0 & $\sim$7000 & (2) & 16 \\ 
W49B & 50308[4,5]010 & 114 & $6663 \pm 1$ & $109 \pm 1$ & 5.0 & 8.0 & 5.8 & 1000--3000 & (2) & 17 \\ 
Cas\,A$^{\ast}$ & 100043020 & 7 & $6617_{-2}^{+3}$ & $435 \pm 9$ & 2.0 & 3.4 & 2.7 & 310--350 & (2) & 18 \\ 
IC\,443 & 5070150[1--4]0 & 368 & $6674_{-13}^{+10}$ & $6.01 \pm 0.59 $ & 0.6 & 1.5 & 10 & 4000--30000 & (3) & 19 \\ 
G292.0+1.8$^{\ast}$ & 506062010 & 44 & $6585_{-28}^{+27}$ & $1.38 \pm 0.35$ & 0.5 & 6.2 & 11 & $\sim$3000 & (3) & 20 \\ 
G349.7+0.2 & 506064010 & 160 & $6617_{-6}^{+7}$ & $2.92 \pm 0.18 $ & 7.0 & 11.5 & 4.0 & $\sim$3500 & (2) & 21,22 \\ 
G350.1--0.3$^{\ast}$ & 506065010 & 70 & $6587_{-10}^{+11}$ & $2.24 \pm 0.23 $ & 3.7 & 4.5 & 1.3 & $\sim$900 & (2) & 23 \\ 
N49$^{\dagger}$ & 807007010 & 185 & $6628_{-26}^{+29}$ & $0.18 \pm 0.04 $ & 0.06 & 50 & 8.5 & $\sim$6600 & (2) & 24 \\ 
N63A & 508071010 & 82 & $6647_{-17}^{+16}$ & $0.86 \pm 0.12 $ & 0.06 & 50 & 10 & 2000--5000 & (2) & 25 \\ 
N132D & (See Note) & 86 & $6656 \pm 9 $ & $1.83 \pm 0.17 $ & 0.06 & 50 & 13 & $\sim$3150 & (2) & 26 \\ 
SN\,1987A$^{\ast}$ & 707020010 & 81 & $6646_{-54}^{+55}$ & $0.19 \pm 0.08 $ & 0.06 & 50 & 0.2 & 27 & (2) & 27 \\ 
\hline
\end{tabular}
\tablecomments{
$^a$The uncertainties are in the 90\% confidence range. 
\\
$^b$The asterisks (${\ast}$) indicate the SNRs for which classification 
is robust from a known association to a compact object, light echo spectroscopy, 
and/or detailed modeling of the ejecta emission. 
The daggers (${\dagger}$) indicate that the progenitor 
type of these SNRs is controversial or unknown.  
\\
$^c$Absorption column density with the solar elemental composition \citep{Wilms00}. 
For the LMC SNRs, only the Galactic component \citep{Dickey90} is shown, 
but the absorption in the LMC ($\lesssim 10^{21}$\,cm$^{-2}$) does not affect the spectra above 5\,keV.
\\
$^d$Background subtraction method we applied (see text in \S2). 
\\
{\bf Observation ID} --- 
RCW\,86: 503004010, 501037010, 503001010, 503002010, 503003010, 503004010 ---
SN\,1006: 500016010, 500017010, 502046010 ---
Sgr\,A East: 100027010, 100037040, 100048010 ---
N132D: 105011010, 106010010, 106010020 
\\ 
{\bf Representative references} --- (1) \cite{Reynolds07}; (2) \cite{Chen99}; (3) \cite{Safi05}; 
(4) \cite{Badenes06}; (5) \cite{Tian11}; (6) \cite{Williams11}; (7) \cite{Yamaguchi08b}; 
(8) \cite{Rakowski06}; (9) \cite{Yamaguchi12b}; (10) \cite{Giacani09}; (11) \cite{Lewis03}; 
(12) \cite{Rest05}; (13) \cite{Warren04}; (14) \cite{Kosenko10}; 
(15) \cite{Koyama07b}; (16) \cite{Koyama07a}; (17) \cite{Keohane07}; (18) \cite{Hwang12}; 
(19) \cite{Troja08}; (20) \cite{Park04}; (21) \cite{Lazendic05}; (22) \cite{Tian14}; (23) \cite{Gaensler08}; 
(24) \cite{Park12}; (25) \cite{Warren03}; (26) \cite{Borkowski07}; (27) \cite{Maggi12} 
}
\end{center}
\end{table*}

We analyzed archival data of all SNRs that {\it Suzaku} has observed to date with 
the X-ray Imaging Spectrometer (XIS), with no bias nor specific selection criterion. 
To search for Fe \Ka\ emission, we extracted XIS spectra from the entire X-ray emitting 
region of each SNR. The only exception was IC\,443, which because of its large angular size 
was only partially imaged by the XIS.  
For this SNR, we extracted the spectrum from a $10'$-diameter circular region in 
the brightest northern part, and estimated the Fe K$\alpha$ flux from the whole SNR 
by scaling the surface brightness using archival {\it XMM-Newton} data. 
Background subtraction was performed in the following manner: 
(1) If nearby blank sky data with taken using an identical detector operating mode were available, 
we used them to extract a background spectrum from the same detector region as the source. 
(2) If the SNR angular size is small enough ($d \lesssim 10'$) compared with the XIS 
field of view, background data were taken from the surrounding region. 
(3) Otherwise, we subtracted only the instrumental background component 
(NXB; simulated by the {\tt xisnxbgen} task), and included models for the extragalactic 
background (a.k.a.\ CXB) and Galactic Ridge X-ray emission (GRXE) in our analysis. 
To estimate the GRXE flux, we followed the relationship between surface brightness 
and Galactic coordinates described by \cite{Uchiyama13}. 
As a consistency check, we applied method (3) to all the SNRs which satisfied the criteria 
for methods (1) and (2), and found no significant change in the measured Fe K$\alpha$ 
blend parameters.

We detected Fe \Ka\ emission from the 23 SNRs listed in Table\,\ref{tab}, 
including five first detections: 
G337.2--0.7, IC\,443, G292.0+1.8, N49, and N63A. 
SNRs without detectable Fe \Ka\ emission can be categorized in two groups: 
evolved SNRs whose electron temperature is too low ($\lesssim 1$\,keV) to 
excite K-shell transitions in Fe atoms, and young SNRs where the hard X-ray 
spectrum is dominated by a strong nonthermal continuum. 
The former category includes most interstellar medium-dominated SNRs 
(e.g., Cygnus Loop, G156.2+5.7, DEM\,L71), while the latter includes 
both shell-like SNRs with cosmic-ray-accelerating blast waves 
(e.g., G1.9+0.3, RX\,J1713.7--3946) and plerionic SNRs (e.g., Crab, G21.5-0.9). 
Recent {\it Chandra} observations of G1.9+0.3 separated spatially the thermal emission
from the nonthermal continuum, and enabled the detection of Fe \Ka\ emission 
\citep{Borkowski10,Borkowski13}. Although {\it Suzaku} observed this SNR for $\sim$100\,ks, 
the Fe \Ka\ emission was not spatially resolved and remained undetected. 
A few SNRs located near the Galactic plane (e.g., G272.2--3.2, Kes\,27) showed hints of 
Fe \Ka\ emission in their NXB-subtracted spectra, but the fluxes were not significantly larger 
than those predicted for the GRXE background, so we excluded them from our study.

We fitted the 5--10\,keV spectrum of each SNR with a power-law 
(or bremsstrahlung) continuum plus a Gaussian for Fe \Ka\ emission. 
Foreground absorption columns are given in Table\,\ref{tab}. 
Some SNRs show emission from Cr, Mn, and/or Ni, and higher transition 
series of Fe K emission, which were modeled using additional Gaussians.  
A radiative recombination continuum (RRC) of Fe\,{\footnotesize XXV} 
was detected in W49B and IC\,443. 
We modeled this component with an exponential function with a threshold energy of 8.8\,keV 
(corresponding to the ionization potential of Fe$^{24+}$, following \cite{Ozawa09b}. 
This is the first detection of the Fe RRC from IC\,443, 
indicating that Fe atoms in this remnant are significantly overionized, 
similar to Si and S \citep{Yamaguchi09}. 
The result is presented in more detail in a separate paper \citep{Ohnishi14}. 
We list the measured centroid energy and unabsorbed flux of the Fe K$\alpha$ blend 
for each SNR in Table\,\ref{tab}.

\section{Discussion}
\label{sec:result}

\begin{figure*}[t!]
  \begin{center}
        \includegraphics[width=18cm]{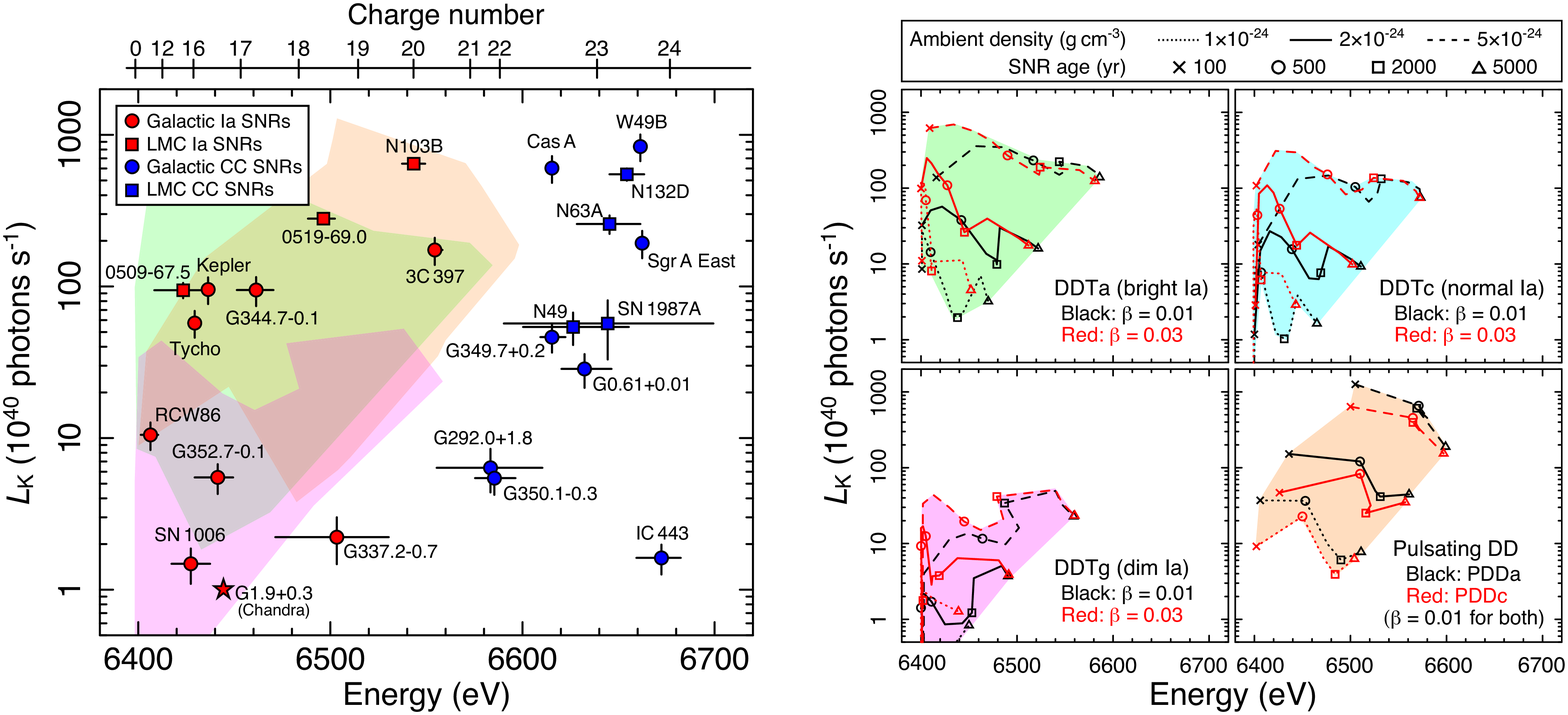}
\caption{\footnotesize
{\it Left}: 
Centroid energies and line luminosities of Fe K$\alpha$ emission from 
various SNRs in our Galaxy (circles) and the LMC (squares). 
The corresponding effective charge number is given above the panel. 
Red and blue represent Ia and CC SNRs or their candidates, respectively. 
The mean centroid and luminosity of Fe \Ka\ emission from G1.9+0.3 observed 
by {\it Chandra} \citep{Borkowski13} are also indicated with the red star. 
The shaded regions represent the Fe \Ka\ centroids and luminosities predicted 
by the theoretical Type Ia SNR models
(DDTa: green, DDTg: magenta, PDD: orange; see right panel for details). \ 
{\it Right}: 
Predicted Fe \Ka\ parameters for the various models of Type Ia SNRs 
evolving in a uniform ambient medium densities of 
$1 \times 10^{-24}$\,g\,cm$^{-3}$ (dotted), 
$2 \times 10^{-24}$\,g\,cm$^{-3}$ (solid),  
and $5 \times 10^{-24}$\,g\,cm$^{-3}$ (dashed).
Crosses, circles, squares and triangles indicate SNR ages of
100\,yr, 500\,yr, 2000\,yr, and 5000\,yr, respectively. 
\label{plot}}
\end{center}
\end{figure*}

Figure\,\ref{plot} shows the Fe K$\alpha$ centroid energy and line luminosity for each SNR, 
together with the corresponding effective charge state $<$$z_{\rm Fe}$$>$ \citep{Yamaguchi14}.  
The line luminosities were calculated from the derived unabsorbed fluxes 
using the distances given in Table\,\ref{tab}.  
The uncertainty in distance to the Galactic SNRs is assumed to be $\pm$10\% 
of the mean values. Although some sources have larger uncertainties 
\citep[e.g., G349.7+0.2, see][]{Tian14}, the fundamental result is not affected, 
since the line centroids play a more important role than the luminosities 
in typing SNRs as described below. 
Also shown in Figure\,\ref{plot} are theoretical predictions for SNR 
models derived from Chandrasekhar-mass Type Ia SN ejecta profiles evolving in a uniform 
ambient medium density in the range of $(1-5) \times 10^{-24}$\,g\,cm$^{-3}$ for SNR 
ages of up to 5000\,yrs \citep[see][for details on the models]{Badenes03,Badenes05,Badenes06}. 
The ejecta profiles shown include three delayed detonation explosion models spanning 
the nominal range of SN Ia kinetic energies and $^{56}$Ni yields (DDTa, DDTc, and DDTg), 
and two pulsating delayed detonation explosions (PDDa and PDDc). 
The predicted Fe K$\alpha$ centroids and luminosities are calculated using updated atomic 
data \citep{Yamaguchi14}. The efficiency of collisionless electron heating at the reverse shock 
is set so that the electron-to-ion temperature ratio is 0.01--0.03 at the immediate post-shock 
region \citep{Badenes05,Yamaguchi14}. 
Similar models for CC SNRs are not available in the literature, but the Ia SNR models and 
the distribution of the data points in Figure\,\ref{plot} allow us to make a number of important 
interpretations using our SNR sample.

First, all {\it bona fide} Ia SNRs have an Fe \Ka\ centroid of $\lesssim$\,6550\,keV 
($<$$z_{\rm Fe}$$>$ $\lesssim$ 20), while all {\it bona fide} CC SNRs have a higher 
centroid energy. 
The separation is very clear: no single object with a robust progenitor type 
(i.e., from a known association to a compact object, or light echo spectroscopy, or 
detailed modeling of the ejecta emission) falls on the wrong side of the centroid boundary. 
Since the Ia and CC SNRs in our sample have similar ages and radii, 
this large difference in the ionization state must be due to significantly higher 
ambient medium densities in the CC SNRs. This is in line with the expectations 
from stellar evolution models, which predict significant (several \Msun) mass loss from 
CC SN progenitors, either due to winds or binary evolution \citep{Langer12}. 
In any case, the clear division in Fe \Ka\ centroid allows us to firmly establish 
the classification of several objects with unclear or controversial types. 
Both RCW\,86 and G344.7--0.1 were once considered to be CC SNRs 
\citep[e.g.,][]{Vink97,Lopez11}, but recent observations have suggested a 
Ia origin \citep[e.g.,][]{Williams11,Yamaguchi12b}. 
The latter is supported by our study. 
The Fe \Ka\ centroid of the Ia SNR G337.2--0.7 \citep{Rakowski06} falls a bit outside 
the range of our theoretical models, but this is not surprising given the estimated age 
(5000--7000\,yr).
\cite{Giacani09} suggested that the highly absorbed SNR G352.7--0.1 might 
have been a CC event, but its Fe \Ka\ centroid puts it squarely in the Ia region. 
The Fe \Ka\ centroid for G1.9+0.3 reported by \cite{Borkowski13} also falls 
within the Ia region, supporting the typing of this SNR by \cite{Borkowski10}. 
Likewise, several SNRs suspected to be of a CC origin without confirmed 
associations with compact objects (e.g., G0.61+0.01: \citealt{Koyama07a}; 
N49: \citealt{Park12}; N63A: \citealt{Warren03}; N132D: \citealt{Borkowski07}) 
fall clearly in the CC region due to their high centroid energies. 
From the Fe \Ka\ centroid alone, the classification for 3C\,397 and N103B is 
somewhat unclear. 
We emphasize, however, that the observed Fe \Ka\ parameters of both SNRs 
can be well reproduced by our Ia SNR models. 
If 3C\,397 is indeed a Type Ia remnant \cite[as suggested by][]{Chen99}, it should 
be relatively old ($\gtrsim 3000$\,yr) and has evolved in a high density interstellar medium 
($\sim 5 \times 10^{-24}$\,g\,cm$^{-3}$), given the comparison with our model plots. 
These values are consistent with the estimates of \cite{Safi05}, 
although a CC origin was suggested in their work. 
The explosion type of N103B has also been a matter of controversy \cite[e.g.,][]{Heyden02,Lewis03}, 
but we favor the Ia hypothesis, based on the high maximum luminosity of 
its parent explosion inferred from the light echo data \citep{Rest05}, 
in addition to the properties of the Fe \Ka\ emission.

Second, within each group, the centroids and line luminosities are fairly well correlated, 
such that SNRs with more highly ionized Fe tend to have more luminous Fe \Ka\ lines. 
This is likely a consequence of the NEI characteristics of the emitting plasma; in order to
collisionally ionize Fe atoms to higher states, higher post-shock densities are required, 
which also result in higher emission measures and thus X-ray luminosities. 
The SNR age also plays a role in the ionization state of the ejecta, but ambient medium 
density is the main driver, as illustrated by the Ia SNR models in Figure\,\ref{plot}. 
Interestingly, both SNR types span a similar range of Fe \Ka\ luminosities, 
despite the fact that typical ejected mass of Fe ($^{56}$Ni) is an order of 
magnitude smaller in CC than in Ia SNe \citep{Woosley95,Iwamoto99}. 
This is again due to the higher emission measure associated with higher post shock density, 
and is one of the reasons why typing SNRs has been difficult without having a detailed model 
for their dynamics and plasma evolution. 
In this context, the relatively high Fe \Ka\ centroid and luminosity for SN\,1987A are 
particularly interesting, given the young age and low ejected Fe mass of this SNR. 
The progenitors of the overionized SNRs, W49B and IC\,443, might have 
an especially high mass loss rate, leading to a CSM dense enough 
to produce a strong circumstellar interaction in the early evolutionary phase 
of their remnants \citep[e.g.,][]{Yamaguchi09}.

Finally, although our Ia SNR models are relatively simplistic (one-dimensional 
hydrodynamics in uniform ambient density), the parameter space 
they span includes {\it all} Ia SNRs. 
This is remarkable in its own right, because some Ia SNRs, 
like Kepler \citep{Reynolds07} and N103B \citep{Lewis03}, 
are known to be interacting with a nonuniform ambient medium. 
Our analysis does not rule out the presence of CSM in these objects, but simply indicates 
that deviations from a uniform ambient medium in Ia SNRs, if present, cannot be very large, 
and rules out large CSM masses (several \Msun) as seen in CC SNRs. 
Middle-aged Ia SNRs with low Fe \Ka\ centroids and luminosities might be interacting with 
an exceptionally low density interstellar medium \citep[e.g., SN1006:][]{Yamaguchi08b}, 
or with a low-density wind-blown cavity excavated by the progenitor 
\citep[e.g., RCW\,86:][]{Williams11}. On the other hand, young Ia SNRs with higher 
Fe \Ka\ centroids and luminosities, like N103B \citep{Lewis03}, 
might be interacting with some kind of CSM, but their Fe \Ka\ emission 
can also be explained by uniform ambient density models, at least at the level of detail 
allowed by our study.

\section{Conclusions}
\label{sec:conclusion}

We have presented a systematic analysis of Fe \Ka\ emission from 23 Galactic and 
LMC SNRs observed by {\it Suzaku}. We find that the Fe \Ka\ line luminosities of 
Type Ia and CC SNRs are distributed in a similar range 
($L_{\rm K}$ = $10^{40-43}$\,photons\,s$^{-1}$), 
but the Fe \Ka\ centroid energies clearly distinguish Ia from CC SNRs, with the former always 
having centroids below $\sim$6.55\,keV and the latter always above. 
We interpret this separation as a signature of different mass-loss rates in Ia and CC SN progenitors. 
The Fe \Ka\ emission of all the Ia objects in our sample is compatible with SNR models 
that expand into a uniform ambient medium, which suggests that Ia progenitors do not modify 
their surroundings as strongly as CC progenitors do. 
This is in line with known limits from prompt X-ray \citep{Hughes07} and radio \citep{Chomiuk12} 
emission from Ia SNe, but our results probe a different regime, constraining the structure of 
the CSM to larger radii (several pc) and progenitor mass loss rates further back in 
the pre-SN evolution of the progenitor. A quantification of these constraints
and a more detailed analysis of the CC SNR sample are left for future work.

The full potential of our method will be realized when it is applied to larger samples of 
higher quality data, as will be accessible to high resolution spectrometers like those on 
{\it Astro-H} and other future missions with large effective areas in the Fe \Ka\ band 
like {\it Athena}. 
These instruments will open the possibility of studying statistically significant samples of 
X-ray emitting SNRs in nearby galaxies with resolved stellar populations like M31, 
which will in turn dramatically increase our knowledge of both Type Ia and CC SN progenitors.

\acknowledgments

We are grateful to Dr.~Katsuji Koyama for providing {\it Suzaku} data he obtained as 
a Principal Investigator and kindly reviewing our manuscript. 
We also thank Drs.\ Kazimierz J.\ Borkowski, Thomas M.\ Dame, Adam, R.\ Foster, 
John D.\ Raymond, Satoru Katsuda, Toshiki Sato, and Hideki Uchiyama 
for helpful discussion and suggestions in preparing this paper.

\bigskip


\end{document}